\documentclass[prl,aps,amsmath,amssymb,amsfonts,twocolumn,nofootinbib,showpacs]{revtex4}
\usepackage{bm,mathrsfs}

\usepackage{graphicx}
\usepackage{epsfig}
\usepackage{amsmath,bbm}
\usepackage{amsfonts,amssymb}
\usepackage{times}
\usepackage{verbatim}

\newcommand{\1}{\mathbbm{1}}

\newcommand{\be}{\begin{equation}}
\newcommand{\ee}{\end{equation}}
\newcommand{\bea}{\begin{eqnarray}}
\newcommand{\eea}{\end{eqnarray}}

\newcommand{\ket}[1]{|#1\rangle}
\newcommand{\bra}[1]{\langle#1|}

\def\>{\rangle}
\def\<{\langle}

\def\qed{\leavevmode\unskip\penalty9999 \hbox{}\nobreak\hfill
     \quad\hbox{\leavevmode  \hbox to.77778em{%
               \hfil\vrule   \vbox to.675em%
               {\hrule width.6em\vfil\hrule}\vrule\hfil}}
     \par\vskip3pt}
     
\begin{document}

\newtheorem{theorem}{Theorem}
\newtheorem{lemma}[theorem]{Lemma}
\newtheorem{corollary}[theorem]{Corollary}
\newtheorem{proposition}[theorem]{Proposition}
\newtheorem{definition}[theorem]{Definition}
\newtheorem{example}[theorem]{Example}
\newtheorem{conjecture}[theorem]{Conjecture}

\title{{ Dissipative preparation of entanglement in optical cavities}}

\author{M. J. Kastoryano$^{1,2}$}
\author{F. Reiter$^{1}$}
\author{A. S. S\o rensen$^{1}$}

\affiliation{$^1$QUANTOP, Danish Quantum Optics Center, Niels Bohr Institute, and\\ $^2$Niels Bohr International Academy, Blegdamsvej 17, DK-2100 Copenhagen \O, Denmark}
\date{\today}

\begin{abstract}  
We propose a novel scheme for the preparation of a maximally entangled state of two atoms in an optical cavity. Starting from an arbitrary initial state, a singlet state is prepared as the unique fixed point of a dissipative quantum dynamical process. In our scheme, cavity decay is no longer undesirable, but plays an integral part in the dynamics. As a result, we get a qualitative improvement in the scaling of the fidelity with the cavity parameters. Our analysis indicates that dissipative state preparation is more than just a new conceptual approach, but can allow for significant improvement as compared to preparation protocols based on coherent unitary dynamics.  
\end{abstract}


\maketitle


Entanglement has for some time now been identified as the key resource in quantum information tasks. In particular, for bipartite systems the properties and uses of maximally entangled states (Bell states) are well understood \cite{Horodecki}. Preparing these Bell states faithfully and reliably has been one of the major challenges in the field of experimental quantum information science, where a plethora of different systems has been investigated \cite{Ladd}. In particular, several schemes based on cavity QED have been proposed (see e.g. \cite{Zoller1,Haroche1,Knight1,Beige,Zheng,SorenMolm,Walther2}), and these schemes have been used to generate entanglement of atoms using microwave cavities \cite{Rauschenbeutel,DiCarlo}.


Traditionally, it has been assumed that noise can only have detrimental effects in quantum information processing. Recently, however, it has been suggested  \cite{DiehlDiss,KrausDiss,DissComp,Beige2}, and realized experimentally \cite{Polzik}, that the environment can be used as a resource. In particular, it was shown in Ref. \cite{DissComp} that universal quantum computation is possible using only dissipation, and that a very large class of states, known as Tensor Product States \cite{MPS,PEPS}, can be prepared efficiently. 
On general grounds, one may argue that dissipative state preparation can have significant advantages over other state preparation methods, but whether it really leads to an advantage for a given task can only be determined by considering concrete physical examples. Below, we argue that this is indeed the case for the system we study. 

\begin{figure}[h!]
\includegraphics[scale=0.35]{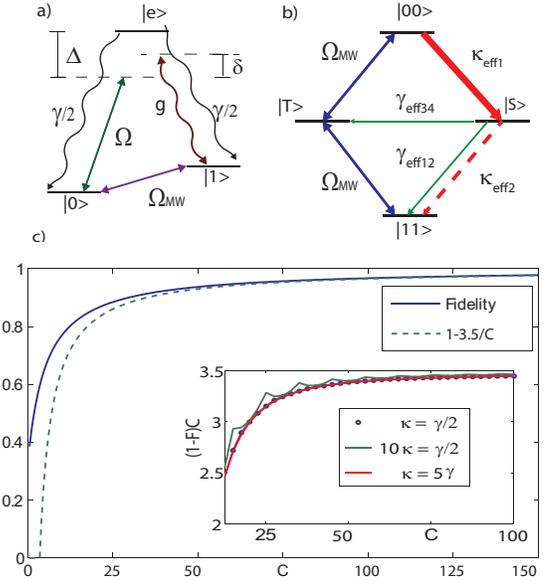}
\caption{\label{fig1abc} (Color online) (a) Level diagram of a single atom with laser detuning $\Delta$ and cavity detuning $\delta$ from two photon resonance. The optical pumping laser for the two atoms differs by a relative phase of $\pi$. (b) The effective two qubit system. The driving $\Omega_{\text{MW}}$ causes rapid transitions between the three triplet states. The decay $\kappa_{\text{eff}}$ takes $\ket{00}$ to $\ket{S}$ and $\ket{S}$ to $\ket{11}$, but as each atom in state $\ket{1}$ creates a shift in the cavity frequency by an amount $g^2/\Delta$, choosing $\delta = g^2/\Delta \gg \kappa$ ensures that only the $\ket{00} \rightarrow \ket{S}$ transition is resonant, because the $\ket{S} \rightarrow \ket{11}$ transition is shifted by twice that amount. This guarantees that $\ket{S}$ is the only stationary state of the dynamical system up to very high fidelity. The spontaneous emission rates $\gamma_{\text{eff},i}$ will tend to reduce the fidelity by redistributing information to the triplet states. (c) Fidelity as a function of the cooperativity $C=g^2/\kappa\gamma$. 
 The inset gives a more accurate account of the scaling ($1-F \leq 3.5 C^{-1}$) for different values of $\kappa/\gamma$.}
\end{figure}

In this letter, we suggest a dissipative scheme for preparing a maximally entangled state of two $\Lambda$-atoms in a high finesse optical cavity \cite{Kimble1,Kimble2,Rempe1}, with detunings as depicted in Fig. 1a. Our scheme can be understood from Fig. 1b, which describes the effective coupled ground states of the atoms in the cavity. A microwave field shuffles the three triplet state around while an effective jump operator describing cavity leakage causes the transitions: $\ket{00} \rightarrow \ket{S} \rightarrow \ket{11}$, where $\ket{S}=(\ket{01}-\ket{10})/\sqrt{2}$ is the singlet state, and $\ket{0}, \ket{1}$ denote logical qubit states. Setting the cavity detuning equal to the single atom cavity line shift will greatly favor the transition to the singlet state and strongly suppress the transition away from it, as the latter is far off resonance; thus driving essentially all of the population into the singlet. 

Our protocol actively exploits the cavity decay to drive the system to a maximally entangled stationary state. The only generic source of noise left in the system is then the one coming from the spontaneous emission. This leads, quite remarkably, to a linear scaling of the fidelity with the cooperativity (see the inset in Fig. 1c), which is in contrast to schemes based on controlled unitary dynamics, where there are two malevolent noise sources, cavity and atomic decay, typically resulting in a weaker square root scaling of the fidelity \cite{Zheng,Walther2,Zoller1,Haroche1,SorenMolm}. In addition to the asymptotic fidelity, one needs to pay particular attention to the time it takes to reach equilibrium. \cite{convcomment}. For our scheme, we show that the convergence rate is rapid. 


We point out that a similar study to ours has been conducted by Wang and Schirmer \cite{WangSchirmer}, where they consider a detuning of the energy levels in order to break the symmetry in the system, and guarantee a unique steady state. It can be shown that their scheme, when generalized to optical cavities, does not give a linear scaling of the fidelity \cite{inprep}, but rather the square root, as for coherent unitary protocols. \\

The basic idea behind dissipative state preparation is to shift the burden of coherence and control from the unitary evolution of the system to the effective noise from the environment. Here we want to engineer the system environment interactions such that the non-unitary quantum dynamical process relaxes to a pure steady state of interest, in our case the singlet state, in a short amount of time. 

In the following, the system-environment interaction will be assumed Markovian, and can thus be modeled by a master equation in Lindblad form:
\be \dot{\rho} = i[\rho,H]+\sum_j L_j \rho L_j^\dag -\frac{1}{2}(L_j^\dag L_j \rho + \rho L_j^\dag L_j), \label{MastEqn}\ee
where the $L_j$'s are the so-called Lindblad operators. We derive a master equation for the ground states of the system, which has the singlet as unique stationary state. This is achieved in our setup, by constructing an effective master equation, whose main contributing terms are the Hamiltonian $H = \frac{1}{2}\Omega_{\text{MW}}(J_+ + J_-)$, and the Lindblad operator $L^{\kappa} = \sqrt{\kappa_{\text{eff}}}\ket{S}\bra{00}$, where $J_+ = \ket{1}\bra{0}\otimes \1 + \1\otimes\ket{1}\bra{0}$.  It can readily be seen that the microwave field ($\Omega_{\text{MW}}$) drives the transitions between the three triplet states ($\{\ket{00},\ket{11},\ket{T}=(\ket{01}+\ket{10})/\sqrt{2}\}$), while the Lindblad operator, originating from the cavity field leakage, will drive the transitions from $\ket{00}$ to $\ket{S}$. The singlet state is thus the unique fixed point of this system, and the relaxation rate can be seen to go as $\min\{\Omega_{\text{MW}}^2/\kappa_{\text{eff}},\kappa_{\text{eff}}\}$, which is just the gap of the Liouvillian if one keeps to the weakly driven and strongly damped regime \cite{gapcomment}. Other terms will contribute weakly to the dynamics of the system, and slightly perturb the stationary state away from $\ket{S}$.\\


We now show how to prepare this effective system in a realistic quantum optical setup.  
Our setup, shown in Fig. 1a, consists of two $\Lambda$-type three level atoms in a detuned cavity with two stable lower energy states $\ket{0}$ and $\ket{1}$, and an excited state $\ket{e}$ with a large energy separation to the lower lying states. We apply one far off-resonance optical laser, with detuning $\Delta$, driving the $0 \leftrightarrow e$ transition and a microwave field  driving the $0 \leftrightarrow 1$ transition resonantly. The cavity mode couples the $1 \leftrightarrow e$ transition off-resonantly, with detuning $\Delta-\delta$, where $\delta$ is the cavity detuning from two photon resonance. Furthermore, we assume a $\pi$ phase difference in the optical laser between the two atoms. This phase difference is crucial in guaranteeing that the singlet is the unique stationary state of the reduced system. Note that one could instead add a phase directly into the coupling term by placing the atoms a half integer cavity wavelengths apart. 

In a rotating frame, this situation is described by the Hamiltonian

\bea H&=& H_0 +H_g + V_+ + V_-\\
H_0 &=& \delta a^\dag a + \Delta(\ket{e}_1\bra{e}+\ket{e}_2\bra{e}) + \notag \\
&&  [ g (\ket{e}_1\bra{1}+\ket{e}_2\bra{1}) a + h.c.]\\
H_g &=& \frac{\Omega_{\text{MW}}}{2} (\ket{1}_1\bra{0} +\ket{1}_2\bra{0})+h.c. \\
V_+&=& \frac{\Omega}{2}(\ket{e}_1\bra{0}-\ket{e}_2\bra{0}), \eea where $V_-=V_+^\dag$, $g$ is the cavity coupling constant, $a$ is the cavity field operator, $\Omega$ represents the optical laser driving strength, and $\Omega_{\text{MW}}$ the microwave driving strength.
On top of the Hamiltonian dynamics, two sources of noise will inherently be present: spontaneous emission of the excited state of the atoms to the lower states with decay rates $\gamma_i$; and cavity leakage at a rate $\kappa$. We assume for convenience that the spontaneous emission rates are the same for decaying to the $\ket{0}$ and to the $\ket{1}$ states (i.e. $\gamma_0 = \gamma_1=\gamma/2$). This translates into five Lindblad operators governing dissipation: $L^\kappa=\sqrt{\kappa}a$, $L^\gamma_{1}=\sqrt{\gamma/2}\ket{0}_1\bra{e}$, $L^\gamma_{2}=\sqrt{\gamma/2}\ket{0}_2\bra{e}$, $L^\gamma_{3}=\sqrt{\gamma/2}\ket{1}_1\bra{e}$, $L^\gamma_{4}=\sqrt{\gamma/2}\ket{1}_2\bra{e}$.

If the optical pumping laser is sufficiently weak, and if the excited states are not initially populated, then the excited states of the atoms, as well as the excited cavity field modes, can be adiabatically eliminated. The resulting effective dynamics will describe two two-level systems in a strongly dissipative environment. To second order in perturbation theory, the dynamics are then given by the effective operators \cite{inprep}:

 \bea H_{\text{eff}} &=& -\frac{1}{2}(V_-H_{NH}^{-1}V_+ + V_-(H_{NH}^{-1})^\dag V_+) + H_g\label{heffective}\\
     L_{\text{eff},j} &=& L_j H_{NH}^{-1} V_+, \label{Leffective}\eea
where $H_{NH} = H_0 -\frac{i}{2}\sum_j L^\dag_j L_j$ is a non-Hermitian Hamiltonian describing the non-unitary dynamics of the excited states which we eliminate. Applying the above equations to our setup, and keeping only terms to lowest order in $\Omega$, the operators in the effective Master equation can be evaluated explicitly, yielding the effective Hamiltonian and principle Lindblad operator

\bea H_{\text{eff}} &=& \frac{1}{2}\Omega_{\text{MW}}(\ket{1}_1\bra{0} + \ket{1}_2\bra{0}+h.c.) + \mathcal{O}(\frac{\Omega^2}{\Delta}) \\
L^\kappa_{\text{eff}} &=& \sqrt{\frac{g^2_{\text{eff}} \kappa/2}{(g^2/\Delta-\delta)^2+(\kappa/2+\gamma\delta/2\Delta)^2}}\ket{S}\bra{00} \label{Lkeff} \\
&&+ \sqrt{\frac{g^2_{\text{eff}} \kappa/2}{(2g^2/\Delta-\delta)^2+(\kappa/2+\gamma\delta/2\Delta)^2}}\ket{11}\bra{S} \notag \eea \label{effL} where $g_{\text{eff}}=g\Omega/\Delta$.

It can be seen from the coefficients in the principal Lindblad operator, that when the cavity detuning is equal to the cavity line shift from a single atom, and when these are large compared to the cavity leakage ($g^2/\Delta=\delta \gg \kappa +\gamma\delta\Delta$), then the denominator of the first term in $L^\kappa_{\text{eff}}$ becomes very small, while the denominator in the second term remains large. The effective Lindblad operator originating from cavity leakage then becomes $L^\kappa_{\text{eff}} \approx g_{\text{eff}} \sqrt{\kappa/2}(1/(\kappa/2+\gamma\delta/2\Delta)\ket{S}\bra{00}+1/\delta\ket{11}\bra{S}) \approx \sqrt{\kappa_{\text{eff}}}\ket{S}\bra{00}$. The effective Hamiltonian shuffles the triplet states among each other, so that the combined effect of the unitary and dissipative dynamics drives essentially all of the population to the singlet state. Hence, we have constructed an effective master equation which approximates the ideal situation described earlier.

We now consider imperfections imposed by spontaneous emission. The four Lindblad operators describing spontaneous emission will also transform into four independent effective noise operators for the reduced system. In the regime discussed above, and keeping only terms which drive the population out of the singlet state, the effective operators for spontaneous emission are 

\bea L^{\gamma}_{\text{eff},i=1,2} &=&  \sqrt{\frac{\gamma^{\text{eff}}}{8}}\ket{11}\bra{S}\notag\\
L^{\gamma}_{\text{eff},i=3,4} &=&  \sqrt{\frac{\gamma^{\text{eff}}}{16}}\ket{T}\bra{S}\label{effGam},\eea where $\gamma_{\text{eff}}=\gamma\Omega^2/2\Delta^2$.

In order to evaluate the performance of the protocol we denote by $P_j$ the probability to be in state $j$, and consider the rate of entering and exiting the singlet state $\dot{P}_S=P_{00}\kappa_{\text{eff},1}-P_{S}(\kappa_{\text{eff},2}+\sum_i \gamma_{\text{eff},i})$, where $\kappa_{\text{eff},1}$ is the square of the first coefficient of $L^\kappa_{\text{eff}}$, $\kappa_{\text{eff},2}$ is the square of the second, and $\gamma_{\text{eff},i}$ are the square of the scalar coefficients of $L^{\gamma}_{\text{eff},i}$. The effective spontaneous emission will tend to modify the steady state population of the three triplet states, but we assume that $\Omega_{\text{MW}}$ is large compared to $\gamma_{\text{eff}}$ and $\kappa_{\text{eff}}$, so that all three triplet states are almost equally populated in steady state. Solving for the stationary state of the rate equation and plugging in the decay rates obtained from the effective operators, we get for $P_S\approx 1$

\be 1-F \approx 3P_{00} \approx 6 \frac{\frac{3}{16}\gamma+\frac{\kappa\Delta^2}{2g^2}}{\frac{g^2\kappa}{2(\kappa/2+\gamma g^2/2\Delta^2)^2}}. \label{scaling1} \ee Here $F$ is the fidelity, which is given by the overlap of the stationary state of the dynamical process with the singlet state $F=|\bra{S}\rho_{ss}\ket{S}|=P_S$. 

There is a trade-off between the second term in the numerator ($\Delta^2\kappa/2g^2$) reflecting the probability to generate a cavity photon by decaying out of the singlet state, and the term in the denominator ($\gamma g^2/2\Delta$) reflecting the scattering of cavity photons off the atoms. The first terms favors a small detuning $\Delta$ to increase the cavity line shift, whereas the second term favors a large detuning $\Delta$ to decrease the scattering. The optimal fidelity is reached when the two terms in the numerator are similar ($\gamma\approx\kappa\Delta^2/2 g^2$), in which case the two terms in the sum in the denominator are also similar ($\kappa\approx\gamma 2 g^2/\Delta^2$). This leads to an error scaling as 
\be 1-F \propto C^{-1}, \label{linScaling}\ee where $C=g^2/\kappa\gamma$ is commonly referred to as the cooperativity.
Plugging in the values in Eq. (\ref{scaling1}) one gets a proportionality factor of roughly 3 in Eq. (\ref{linScaling}). By numerically extracting the fixed point of the full master equation, and then maximizing its fidelity with respect to the singlet for fixed values of $C$, we get that the actual constant is closer to 3.5 (inset in Fig. 1c),  i.e. $1-F \approx 3.5 C^{-1}$. This discrepancy can be attributed to the fact that we did not include all of the spontaneous emission terms in Eq. (\ref{scaling1}). In addition, the assumption that all three triplet states are equally populated is not exact. To support our analysis further, we also note that the fidelity scaling is essentially independent of the ratio $\kappa/\gamma$ (inset in Fig. 1c).\\ 

\begin{figure}[t]
\includegraphics[scale=0.35]{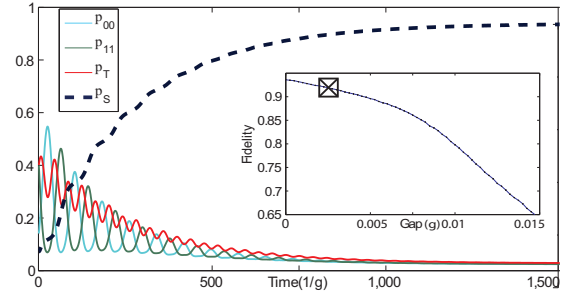}
\centering
\caption{\label{fig2Gap} (Color online) The main figure shows the population of the singlet state (thick dashed line) and of the three triplet states (full lines) as a function of time for a random initial state. The curves were plotted for $C=50$, $\kappa=\gamma/2$, $\Omega=5\Omega_{\text{MW}}/2$, $g=20\Omega$, and $\Delta$, $\delta$ are such that they maximize the fidelity for small $\Omega$. In this parameter regime, the stationary state has a $92\%$ fidelity with respect to the singlet state. The inset shows the maximal fidelity as a function of the gap size for $C=50$ and $\kappa=\gamma/2$. The main figure corresponds to the cross on the curve in the inset.}
\end{figure}

For comparison, in a controlled unitary dynamics protocol, the fidelity will suffer errors coming from spontaneous emission on the one hand, and from cavity decay on the other. Decreasing one of the error sources will typically increase the other in such a way that the optimal value of the fidelity is $1-F \propto 1/\sqrt{C}$ \cite{SorenMolm}. Indeed, to the best of our knowledge, all entangled state preparation protocols based solely on controlled unitary dynamics scale at best as $1/\sqrt{C}$ \cite{Zheng,Zoller1,Haroche1,Walther2}. This means that the linear scaling of the fidelity from Eq. (\ref{linScaling}), is a quadratic improvement as compared to any known closed system entanglement preparation protocol. We note, however, that it is possible to beat this if one exploits measurement and feedback \cite{Beige,Knight1,SorenMolm}. As mentioned previously, the reason for this improvement stems from the fact that cavity decay is used as a resource in our dissipative scheme, so that the only purely detrimental source of noise is the spontaneous emission.

The above analysis has been conducted without any consideration of the speed of convergence. We now show, that the entangled stationary state can be reached rapidly. In Fig. 2, we simulate the dynamics of the full master equation for an appropriate set of parameters. Starting from an arbitrary initial state, the populations of triplet states undergo rapid coherent oscillations with an envelope decaying at a rate proportional to the gap \cite{gapcomment}, while the singlet state converges to its maximum value at the same rate. 

For a given cavity, $g$, $\kappa$, and $\gamma$ can be considered fixed by experimental constraints, and the speed of convergence is primarily governed by the magnitudes of $\Omega$ and $\Omega_{\text{MW}}$. The speed of convergence can be increased by increasing the driving laser strength ($\Omega$), but the latter can not be too large otherwise perturbation theory breaks down, and the excited cavity and atomic states can no longer be ignored. Furthermore, $\Omega_{\text{MW}}$ can not be too small with respect to $\{\kappa_{\text{eff}},\gamma_{\text{eff}}\}$, otherwise the coherent shuffling of the triplet states will not be sufficiently strong to keep them at equal population. The inset in Fig. 2 shows how the maximal fidelity scales as a function of the gap for a specific set of cavity parameters. The curve is plotted by optimizing the fidelity, for given fixed values of the gap, with respect to $\{\Omega, \Omega_{\text{MW}}\}$, for fixed values of $\{\Delta,\delta\}$ (those which are optimal for small $\Omega$). There is clearly a trade-off between the accuracy of the dissipative state preparation protocol and the speed at which one reaches the stationary state, but close to optimal fidelity the dependence is weak. 

Present day experimentally achievable values for the cooperativity are around $C \approx 30$ \cite{Rempe1,Kimble1,Kimble2}. This puts our scheme at $\sim 90\%$ fidelity with respect to the singlet state. Fig. 2 shows that the stationary state is reached in a time $\sim 1000/g$, which yields for $g=(2\pi)35$ MHz \cite{Kimble2} a convergence time of roughly $5$ $\mu$s starting from an arbitrary initial state. \\

We have proposed a scheme for dissipatively preparing an entangled state of two trapped atoms in an optical cavity. From both analytical and numerical evidence, we give the scaling of the error explicitly, and show that the stationary state is reached rapidly. Our results indicate that not only can one produce entanglement dissipatively in a simple cavity system, but, to the best of our knowledge, the scaling of the fidelity for such entanglement preparation is better than any existing coherent unitary protocol. These results are an indication that an approach based on dissipation can be very fruitful for state preparation, as one manifestly can transform a previously undesirable noise source into a resource. It would be interesting to see if one could obtain similar results in related systems such as trapped ions and solid state based quantum devices, where dissipation traditionally plays a detrimental role. 

We thank M.M. Wolf and D. Witthaut for helpful discussions. We acknowledge financial support from the European project QUEVADIS and from the Villum Kann Rasmussen Foundation.


\begin{thebibliography}{24}
\bibitem{convcomment} For larger systems (typically lattice systems), although the fixed point of the dynamical process might be the desired state, if the convergence is too slow then the scheme cannot be viable in practice.
\bibitem{gapcomment} The decay rate of the quantum dynamical process is often governed by the spectral gap of the Liouvillian, i.e. the magnitude of the difference between the first and second largest real parts of the eigenvalues. 
\end{thebibliography}
\end{document}